\documentclass[conference]{IEEEtran}
\IEEEoverridecommandlockouts

\usepackage{cite}
\usepackage{amsmath,amssymb,amsfonts}
\usepackage{algorithmic}
\usepackage{graphicx}
\usepackage{textcomp}
\usepackage{url}
\usepackage{minted} 
\usepackage{xcolor}
\definecolor{bg}{rgb}{0.95,0.95,0.95} 
\usepackage{svg}
\usepackage[hidelinks]{hyperref}
\begin{document}

\title{Describing Agentic AI Systems with C4:\\ Lessons from Industry Projects}

\author{

\IEEEauthorblockN{Andreas Rausch}
\IEEEauthorblockA{\textit{Institute for Software and Systems Engineering} \\
\textit{Clausthal University of Technology}\\
Clausthal-Zellerfeld, Germany \\
arau@tu-clausthal.de}

\and

\IEEEauthorblockN{Stefan Wittek}
\IEEEauthorblockA{\textit{Institute for Software and Systems Engineering} \\
\textit{Clausthal University of Technology}\\
Clausthal-Zellerfeld, Germany \\
switt@tu-clausthal.de}

}

\maketitle

\begin{abstract}
Different domains foster different architectural styles---and thus different documentation practices (e.g., state-based models for behavioral control vs.\ ER-style models for information structures). Agentic AI systems exhibit another characteristic style: specialized agents collaborate by exchanging artifacts, invoking external tools, and coordinating via recurring interaction patterns and quality gates. As these systems evolve into long-lived industrial solutions, documentation must capture these style-defining concerns rather than relying on ad-hoc code sketches or pipeline drawings.

This paper reports industrial experience from joint projects and derives a documentation systematics tailored to this style. Concretely, we provide (i) a style-oriented modeling vocabulary and a small set of views for agents, artifacts, tools, and their coordination patterns, (ii) a hierarchical description technique aligned with C4 to structure these views across abstraction levels, and (iii) industrial examples with lessons learned that demonstrate how the approach yields transparent, maintainable architecture documentation supporting sustained evolution.

\end{abstract}


\section{Introduction}
\label{sec:introduction}

Major technology shifts yield new architectural styles and, eventually, fitting description techniques. Different domains foster different styles---and thus different documentation practices (e.g., state-based models for behavioral control vs.\ ER-style models for information structures). For instance, the Internet/REST era established characteristic web-system architectures \cite{fielding2000rest} along with recurring description elements such as clients, services, resources identified via URIs, HTTP methods, and API contracts (e.g., OpenAPI specifications) \cite{openapi}. Likewise, ROS~2 shaped recurring architectures for distributed robotic systems \cite{macenski2022ros2} and domain-specific description elements (nodes, topics, services, launch configurations). Architectural description evolves accordingly to support communication, maintainability, and systematic evolution.

A comparable shift is currently underway with \emph{LLMs} and \emph{agentic AI frameworks}. Many organizations are moving from ``AI as a component'' to \emph{agentic AI systems} in which specialized agents collaborate, interact with users, and invoke external tools (e.g., tool/function calling, MCP-style tool access) to automate tasks that were previously performed manually. Typical examples include generating executable test scripts from requirements or estimating the price of a used product on the resell market. 

In our joint projects with industrial partners, we observed that these agentic AI systems quickly become long-lived, evolving products. They accumulate dependencies and constraints around agent responsibilities and boundaries, interaction protocols, exchanged artifacts and memory, tool interfaces, and operational governance (permissions, budgets, validation, human approvals). Yet in practice, many implementations remain documented in an ad-hoc manner---often as informal pipeline sketches or code-level structures---making change-impact analysis, maintenance, and systematic evolution unnecessarily difficult.

Architecture description for \emph{agentic AI systems} in industrial contexts is still emerging. While individual modeling approaches exist, they are rarely integrated into established architecture documentation frameworks such as C4 \cite{brown_c4} or arc42 \cite{arc42}. To document agentic AI systems within such frameworks, the documentation must reflect their characteristic style---agents collaborating via artifacts, tool invocations, and coordination patterns with quality gates. This motivates a lightweight systematics that integrates these style-specific concerns into a hierarchical, framework-compatible documentation structure.

This paper is an \emph{industry experience report} that reflects on what has worked for us in practice when describing agentic AI systems. Rather than proposing a comprehensive new ADL, we provide (i) an architectural style centered on \emph{agents, artifacts, tools, and users} and their interaction patterns, (ii) a \emph{hierarchical, decomposition-based functional description technique} aligned with established architecture documentation approaches such as C4, and (iii) representative industrial examples with cross-case lessons learned. Together, these results support understanding, maintenance, and systematic evolution of agentic AI systems.

The remainder of this paper is structured as follows: Section~\ref{sec:related-work} reviews related work on architectural style anchors and lightweight documentation practices. Section~\ref{sec:approach} details the architectural style observed in our projects and presents our hierarchical description technique aligned with C4. Section~\ref{sec:industrial-examples} demonstrates the approach through tree industrial examples---a requirements-to-test-script generator, a blueprint-guided architecture recovery system and a resell price estimation system. Additional the section also consolidates lessons learned. Finally, Section~\ref{sec:conclusion} concludes the paper and outlines directions for future work.

\section{Related Work}
\label{sec:related-work}

\subsection{Architectural Style Anchors for Agentic AI Systems}
\label{sec:related-styles}
Agentic AI systems combine multiple specialized agents, tool access, and human interaction in recurring architectural structures.
From a software-architecture perspective, many realizations resemble variants of established styles:
(i) \emph{pipes-and-filters} for staged processing pipelines \cite{ms_pipesfilters},
(ii) \emph{blackboard} architectures that coordinate specialized problem solvers via a shared workspace \cite{corkill1991blackboard}, and
(iii) \emph{orchestration vs.\ choreography} as alternative collaboration patterns for distributed components \cite{megargel2021orchchor}.
These anchors provide useful intuitions for decomposition and interaction, but they do not explicitly address agentic-specific concerns such as bounded autonomy, artifact-centric handoffs, and operational quality gates.

Closer to our scope, research on multi-agent systems (MAS) has long captured coordination knowledge in pattern form, including systematic overviews of MAS design patterns \cite{juziuk2014maspatterns}.
More recently, Liu et al.\ present an \emph{agent design pattern catalogue} for foundation-model-based agents \cite{liu2025agentpatterns}.
While such catalogues are valuable for \emph{designing} agentic systems, they do not yet provide a lightweight, practitioner-friendly \emph{architecture description} approach that supports communication, maintenance, and evolution across industrial teams.

\subsection{Architecture Description and Agentic Modeling Practices}
\label{sec:related-ad}
Architecture description standards and practitioner approaches provide complementary guidance on \emph{how to document} architectures. ISO/IEC/IEEE~42010 structures architecture descriptions around stakeholders, concerns, viewpoints, and views \cite{iso42010_2022}. For pragmatic day-to-day documentation, the C4 model communicates software architectures via a small set of scalable views \cite{brown_c4}, and arc42 provides a widely used, practitioner-oriented template for architecture documentation \cite{arc42}. However, these frameworks are largely technology-agnostic and do not prescribe agentic-specific documentation constructs (e.g., explicit agent roles and boundaries, interaction protocols/workflows, artifact and memory structures with provenance, tool access, and operational quality gates).

In parallel, first modeling and reference-architecture approaches tailored to agentic AI are emerging. For example, Ait et al.\ propose a BPMN extension for modeling human--agentic collaborative workflows \cite{ait2025bpmn}. Beyond workflow modeling, several works propose agentic-specific architectural guidance, including reference architectures \cite{lu2024refarch}, architecture option taxonomies and decision models \cite{zhou2024taxonomy}, and architecture proposals for LLM-based multi-agent systems \cite{becattini2025sallma}. Complementary efforts capture recurring agent structures and coordination knowledge in catalogs and patterns \cite{liu2025agentpatterns} or provide unified modeling frameworks for designing agent architectures \cite{hassouna2026llmagentumf}. While valuable, these approaches are typically presented in isolation and are not yet integrated into holistic, hierarchical architecture documentation frameworks such as C4 or arc42, limiting their use as a consistent systematics for communication, maintenance, and evolution in industrial practice.

\section{Architectural Style and C4-Based Hierachical Description for Agentic AI Systems}
\label{sec:approach}

Across multiple joint industry projects, we observed that successful agentic AI systems consistently adopt a recognizable architectural pattern: a \emph{group of specialized agents} collaborates by exchanging \emph{artifacts}, invoking \emph{internal and external tools}, and coordinating decisions through explicit interaction patterns. This aligns with emerging catalogs for foundation-model-based agents, which describe these systems as cooperating ensembles with recurring coordination mechanisms and shared work artifacts \cite{liu2025agentpatterns}.

In practice, two complementary coordination modes predominate: \emph{orchestration}, where a dedicated conductor delegates tasks to worker agents, and \emph{choreography}, where peer agents react to events, such as artifact updates, to trigger subsequent processing steps \cite{megargel2021orchchor}. A recurring structural nucleus is the \emph{planner--executor} pattern, often supplemented by a \emph{verifier/critic} for validation \cite{liu2025agentpatterns}. While orchestrators impose a collaboration protocol (defining actor sequences and quality gates), choreographed variants realize these protocols via event-driven reactions. Importantly, these architectures are typically \emph{hierarchical}; an agent ensemble can be treated as a higher-level subsystem, necessitating a description technique that supports multi-level decomposition rather than flat workflow representations.

To document this style throughout the system lifecycle, we propose a hierarchical decomposition aligned with the C4 model. This approach positions the agentic system within its \emph{Context} and decomposes it down to the \emph{Component} and \emph{Code} levels, making ensembles, individual agents, and their respective interfaces explicit. This multi-level approach is essential for maintaining concise descriptions while enabling change-impact analysis and evolution planning.

\begin{figure*}[htbp]
    \centering
    \includegraphics[width=0.87\linewidth]{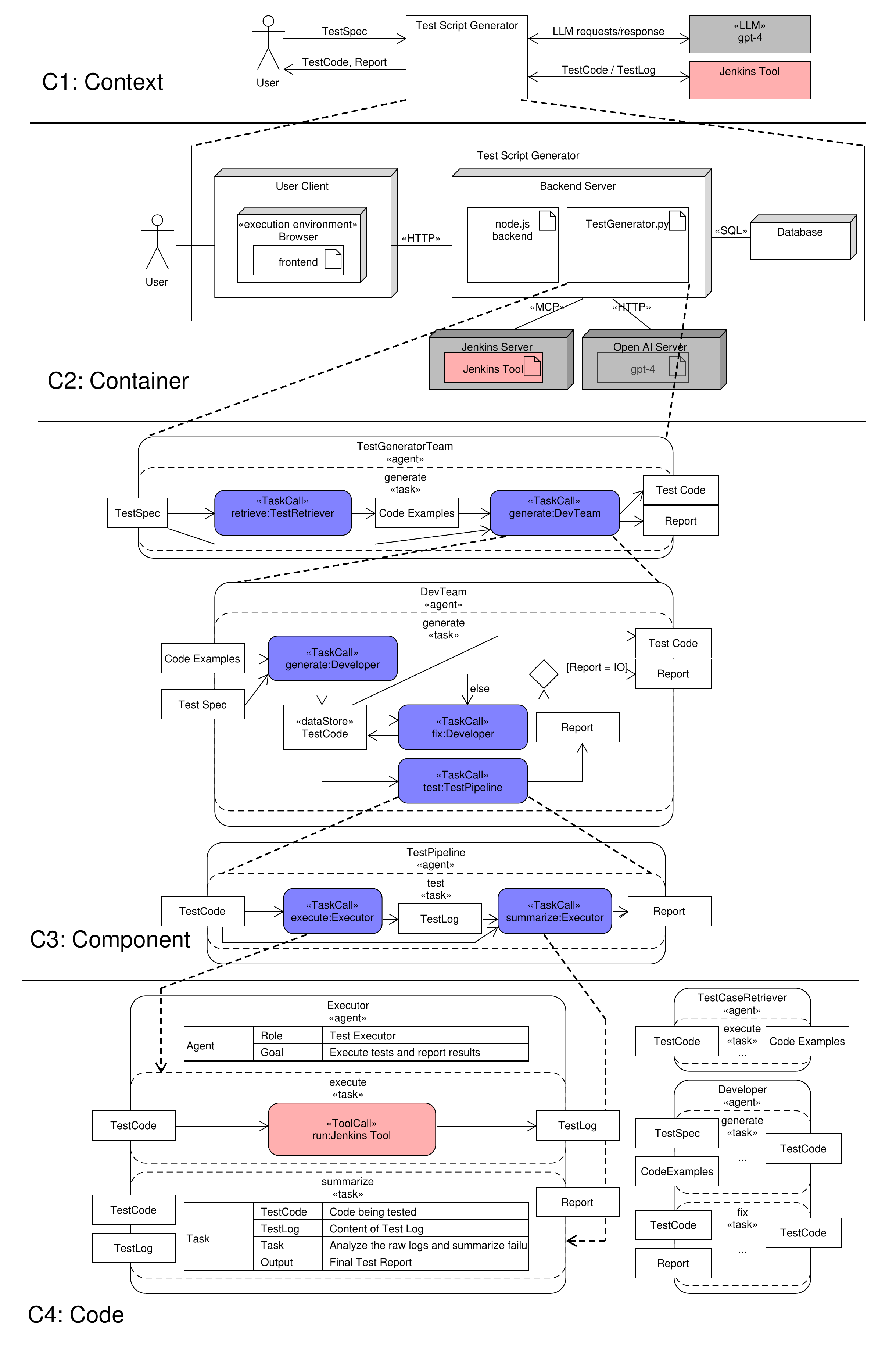}
    \caption{All C4 architectural levels of Test Script Generator example.}
    \label{fig:c4 example}
\end{figure*}

Figure \ref{fig:c4 example} illustrates our proposed notation and architectural description for the Test Script Generator System. This diagram encompasses all four levels of the C4 model, adapted for agentic systems. 

\subsection{C1: Context}
The \emph{C1 Context level} provides a high-level overview of the system, depicting the system itself, the Users, and any external Systems. Connections at this level are directional and explicitly labeled with the artifacts being exchanged. A distinguishing feature of our notation is the explicit depiction of the Large Language Model (LLM) driving the agents; this allows for the specification of particular LLM versions or the use of multiple models. Additionally, all external tools are visually highlighted with a red background.

\subsection{C2: Container}
The \emph{C2 Container} level zooms into the system boundary to provide a runtime view. We utilize UML Deployment Diagram syntax to depict the system nodes. Connections at this level represent specific protocols, such as HTTP or the Model Context Protocol (MCP). Consistent with the previous level, external tools remain distinctively marked with a red background.

\subsection{C3: Component}

The C3 level provides a hierarchical decomposition of the systems behavior. It is based on UML Activity Diagram syntax to describe argentic ai specific elements.

\subsubsection{Agents}
Are activities with the stereotype \texttt{<<agent>>} represent single agents (e.g. \emph{Executor} later described in C4), groups of agents, and their workflows (e.g. \emph{TestPipeline}). 
\subsubsection{Task}
To visually bundle behavior into specific tasks (e.g., \emph{test}), we utilize \emph{interruption activity regions}—represented as rounded boxes with dashed lines—marked with the stereotype \texttt{<<task>>}.
\subsubsection{Artifacts}
Object nodes are used to specify artifacts, defining both inputs (e.g., \emph{TestCode}) and outputs (e.g., \emph{Report}). Crucially, in this notation, it is the Tasks that require inputs and produce outputs, not the Agents themselves, as these requirements vary based on the specific task being performed. An agent may be responsible for multiple tasks.
\subsubsection{Memory}
Object nodes with the \texttt{<<datastore>>} stereotype are used to model a shared memory (i.e. for \emph{TestCode} in the \emph{DevTeam}) that agents update during execution.  
\subsubsection{Quality Gates}
Agent decisions can be modeled using Decision Node and Guards (i.e. the quality gate [Report=IO] in generate:DevTeam, this is an instance of the \emph{verifier/critic pattern}). 
\subsubsection{TaskCall}
A central addition to our notation is the Call Behavior Action with the stereotype \texttt{<<TaskCall>>}. As the name suggests, a \texttt{<<TaskCall>>} realizes the invocation of a specific task on a specific agent (e.g., \emph{execute:Executor}). This facilitates our suggested decomposition strategy: Task Calls are decomposed into Agent Tasks, which are further composed of Task Calls etc.
\subsubsection{Interaction Pattern}
Our notation uses activity diagram syntax to model the interaction patterns of the agents. \emph{DevTeam} exampels a workflow like interaction, here the agents produce artifacts in a defined pipeline combined with a quality gate and loops. The Resell App described later follows the orchestration interaction pattern.

\subsection{C4: Code}
The \emph{C4 Code} level represents the lowest level of decomposition, consisting of leaf actions. While it retains the C3 syntax, it introduces leaf actions to describe implementation details. 
\subsubsection{Tools} 
The \emph{ToolCall} element depicts the invocation of internal or external tools (e.g., \emph{run:JenkinsTool}).
\subsubsection{Prompts}
Furthermore, this level includes specific descriptions of the Prompts passed to the LLM (e.g., in a \emph{summarize:Extractor} Task), presented as a table. The structure typically follows the specific Agentic Framework used, distinguishing between static elements (such as role descriptions) and task-specific elements (such as current task descriptions and input artifacts like \emph{TestCode} or \emph{TestLog}).

\section{Industrial Examples}
\label{sec:industrial-examples}
\subsection{Example - Test Script Generator}
\label{sec:example-a}

The primary objective of the Test Script Generator is to automate the creation of GUI test scripts based on formal Requirement Specifications. This system is specifically designed to automate the currently manual testing efforts of a large legacy system (System Under Test). To enhance the quality and syntactic correctness of the output, the generation process is augmented by a knowledge base of existing test scripts, which serve as reference examples (implemented in \emph{retrieve:TestRetriever}).

Figure \ref{fig:c4 example} depicts a selection of the architectural description, illustrating the system across all levels of the C4 model. A focal point of this architecture is the Core Generation Loop defined at the C3 Component level. This workflow is orchestrated by the \emph{GeneratorTeam}.

The process initiates by ingesting retrieved code examples (\emph{Code Examples}) and the target Test Specification (\emph{TestSpec}). The \emph{generate:Developer} task utilizes these inputs to synthesize an initial version of the test code. Subsequently, the \emph{test:TestPipeline} executes this code against the legacy system, producing a detailed execution report.

A critical feature of this architecture is its self-correcting capability. If the execution results in errors (classified as "Not In Order" or NIO), an iterative feedback loop is triggered. The \emph{fix:Developer} task analyzes the flawed code alongside the error report to generate a corrected version. Note that this update mechanism is described using the dataStorage of UML Activity Diagrams. This cycle repeats until the test executes successfully. Finally, the validated \emph{TestCode} and the corresponding \emph{Report} are released to the wider system as final artifacts.

\subsection{Example - Blueprint-Guided Architecture Recovery System}
\label{sec:example-b}
\begin{figure}[htbp]
    \centering
    \includegraphics[width=1\linewidth]{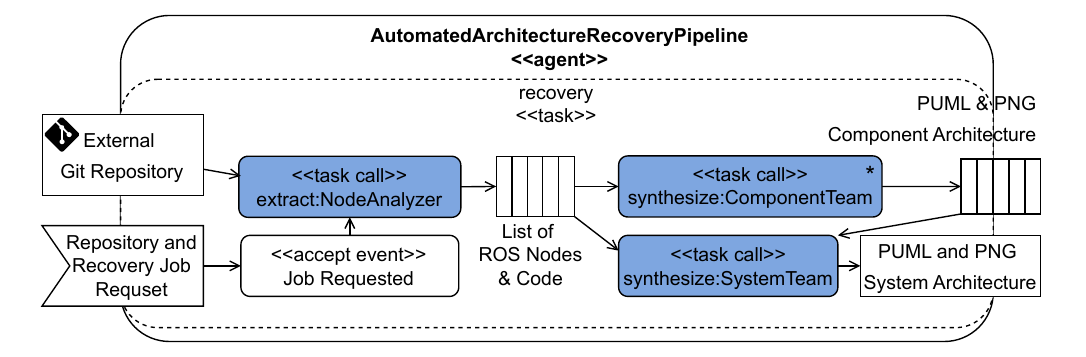}
    \caption{Part of the \emph{C3: Component} level description of \emph{Blueprint-Guided Architecture Recovery} example.}
    \label{fig:Copilot-Architecure}
\end{figure}
As a second case study, we introduce the \emph{Blueprint-Guided Architecture Recovery System}. In contrast to the generation use case, this system focuses on the reverse engineering of architectural models from existing source code. Specifically, it targets ROS~2 systems, aiming to recover hierarchical functional architectures by analyzing code and launch configurations. The recovery process is guided by a formal notation defined in an architectural blueprint, ensuring that the inferred models conform to specific domain constraints.

Figure \ref{fig:Copilot-Architecure} illustrates the core workflow of this system using our C3 notation. The entire process is encapsulated within the \emph{AutomatedArchitectureRecoveryPipeline}. The workflow is initiated by an external \emph{Job Request} (control flow) and operates on an external \emph{Git Repository} (object flow), demonstrating a clear separation between execution triggers and data sources.

Following the deterministic extraction phase, the workflow transitions into the generative stage. Here, the \emph{List ROS Nodes \& Code} serves as the central interface artifact. The processing of this list showcases the expressive power of our activity-diagram-based syntax:
\begin{itemize}
    \item The segmented object node depicts the list as a collection of elements.
    \item The asterisk (*) atop the \emph{synthesize:ComponentTeam} \texttt{<<TaskCall>>} indicates that the list is processed element-wise, producing a distinct \emph{PUML \& PNG of Component Architecture} for each identified node.
    \item Conversely, the list is processed as a holistic entity by the \emph{synthesize:SystemTeam} \texttt{<<TaskCall>>} to reconstruct the global system-level hierarchy and composition, which is finally outputted.
\end{itemize}

\subsection{Example - Resell App}
\label{sec:example-c}
\begin{figure}[htbp]
    \centering
    \includegraphics[width=0.7\linewidth]{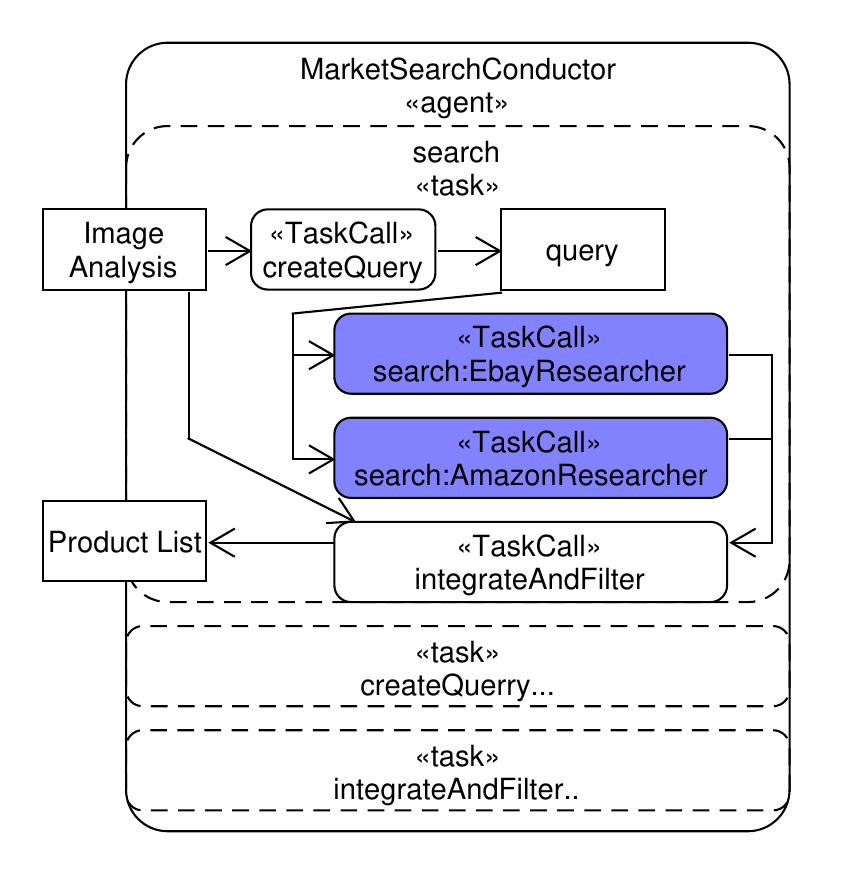}
    \caption{Part of the \emph{C3: Component} level description of the \emph{Resell App System} example.}
    \label{fig:ResellApp}
\end{figure}

The Resell App streamlines selling used products on platforms like eBay. It uses a Multimodal Large Language Model to analyze images, followed by a MarketSearchConductor agent that identifies matching offers to estimate prices. Figure \ref{fig:ResellApp} illustrates this agent in C3 Notation. First the agent derives a query from the image Analysis. This query is than delegated to a Ebay and a Amazon researcher. Both provide a Product List, which is integrated and filed toward relevance with respect to the original image analysis. Note that this is an instance of the orchestration interaction pattern.

Note also that in this case the agent is also calling tasks on itself (\emph{createQuery} and \emph{integrateAndFilter}). To make this clear in the notation, we drop the instance marking (\emph{:MarketSearchConductor}) and the color. 

\subsection{Lessons Learned}
The application of our adapted C4 notation to agentic systems has revealed several insights regarding the balance between formal modeling and architectural clarity.
A primary observation concerns the \emph{C2 Container Level}, where we noted a loss of information regarding artifacts passed over connections. Because standard deployment diagrams prioritize physical node distribution and communication protocols, such as HTTP or MCP, they do not natively represent internal interfaces or artifact exchanges between components on the same node. While prioritizing protocols maintains a clean runtime view, it is suboptimal as these artifacts inevitably reappear in detailed C3 and C4 decompositions.

A further challenge is the redundancy of input object nodes required to maintain task interface transparency. Artifacts like \emph{TestCode} must often be duplicated within an agent's task region to clarify specific inputs for tasks. While this repetition does not strictly adhere to standard UML activity diagram syntax, it is necessary to ensure that the interface requirements of discrete tasks remain explicit.

Finally, we observed that iconization could significantly enhances readability. Our current notation relies heavily on text-based stereotypes like \texttt{<<agent>>} and \texttt{<<task>>}, which increase cognitive load as system complexity grows. By incorporating domain-specific icons, we can reduce visual clutter and improve the immediate understanding of the system. 

\section{Conclusion and Future Work}
\label{sec:conclusion}
This paper introduced an adapted C4-adherent, UML-based notation for architecting agentic AI systems, demonstrated through a test script generation loop,  a blueprint-guided ROS~2 architecture recovery pipeline and a resell price estimation system.

Future work should focus on refining the C2 level to better bridge the gap between deployment structure and functional data flow and establishing a standardized icon set that maintains formal UML semantics while providing a more intuitive experience for architects. Additional an explicated notation for different interaction patterns like orchestration or choreography would be a promising direction.

Beyond immediate visualization benefits, this work contributes to the emerging discipline of \emph{AI Engineering}. By treating agentic workflows as first-class architectural concerns—rather than opaque implementation details—organizations can enforce governance, auditability, and systematic reuse. As these systems scale, the ability to decompose complex agent ensembles into understandable, reviewable views will be the deciding factor between fragile prototypes and robust industrial products. 

\bibliographystyle{IEEEtran}
\bibliography{IEEEabrv,main}

\end{document}